# Investigating the Timing Behavior of Compton Scattering in BGO for Time-of-Flight PET

Minseok Yi[1,2,3], Daehee Lee[3], Alberto Gola[4], Stefano Merzi[4], Michele Penna[4], Simon R. Cherry[3], Jae Sung Lee[1,2,5,*], and Sun Il Kwon[3,*]

[1]Interdisciplinary Program in Bioengineering, College of Engineering, Seoul National University, Seoul 03080, Korea
[2]Integrated Major in Innovative Medical Science, Seoul National University, Seoul 03080, Korea
[3]Department of Biomedical Engineering, University of California, Davis, One Shields Avenue, Davis, CA 95616, USA
[4]Fondazione Bruno Kessler, via Sommarive 18, Trento I-38123, Italy
[5]Brightonix Imaging Inc., Seoul 04782, Korea

E-mail: sunkwon@ucdavis.edu & jaes@snu.ac.kr

## Abstract

**Bismuth germanate (BGO) is gaining renewed attention as a viable material for hybrid Cherenkov/scintillation time-of-flight positron emission tomography (TOF-PET) detectors. While single-crystal studies have demonstrated excellent timing resolution by leveraging prompt Cherenkov photons, practical detector modules based on pixelated arrays introduce a high prevalence of inter-crystal scattering (*InterCS*) events, complicating timing accuracy. In this study, we experimentally investigated the impact of *InterCS* on BGO Cherenkov timing using a dual-pixel detector coupled to a segmented SiPM readout. Events were classified into full-energy deposition (*FED*; primary crystal 511 keV absorption), *InterCS*, and penetration types via energy-weighted positioning and validated using GATE simulations, which also revealed that over 25% of the experimentally identified full-energy events involved intra-crystal scatter (*IntraCS*). For *InterCS* events, the optimal timestamp selection was achieved by choosing the earlier of the two timestamps, yielding a coincidence timing resolution of 221 ps FWHM (831 ps FWTM)—markedly worse than the 184 ps (603 ps FWTM) obtained for *FED* events. Furthermore, prompt photon yield was found to decrease measurably due to energy splitting: *InterCS* events averaged 4.73 detected photons in the first 1 ns, compared to 5.76 for *FED* events. These results emphasize the importance of incorporating time-aware, per-pixel timestamping strategies in pixelated BGO TOF-PET systems to maintain optimal timing performance in the presence of scatter.**

## 1. Introduction

Bismuth germanate (BGO) has recently gained attention as a candidate material for hybrid scintillator/Cherenkov time-of-flight positron emission tomography (TOF-PET) detectors. Compared to LSO and LYSO, BGO offers higher stopping power, lower cost and no significant intrinsic radioactivity, making it attractive for high-sensitivity, cost-effective PET imaging **[1-7]**. However, its inherently slow scintillation decay limits the timing performance of current commercial PET scanners based on BGO.

To overcome this limitation, recent studies have focused on utilizing the promptly emitted Cherenkov photons in BGO to achieve fast timing **[8, 9]**. Although the Cherenkov photon yield in BGO is low— approximately 17 Cherenkov photons are produced per 511 keV event—these photons arrive within tens of picoseconds of the gamma interaction, enabling high-resolution time pickoff when efficiently detected. Prior studies have demonstrated promising coincidence timing resolution (CTR) using single-pixel BGO detectors with optimized readout, triggering, and signal processing strategies **[10-18]**.

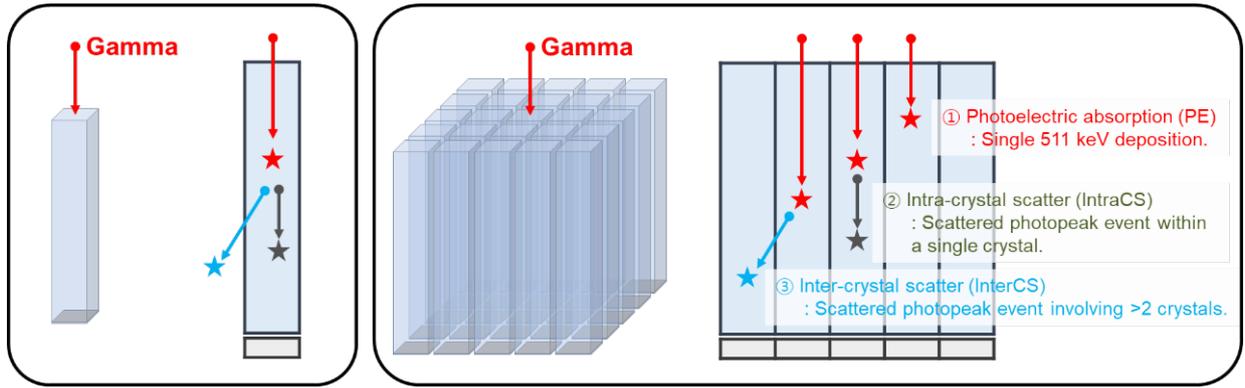

**Figure 1. Conceptual illustration of event types in pixelated BGO Cherenkov TOF-PET detectors.** In contrast to single-crystal setups, pixelated detector configurations introduce additional complexities due to gamma-ray scatter between crystals. We focus on three categories: (1) pure photoelectric absorption (*PE*), where all energy is deposited at a single site; (2) intra-crystal scatter (*IntraCS*), where multiple energy depositions occur within the same crystal; and (3) inter-crystal scatter (*InterCS*), where energy is shared across neighbouring pixels. Since Cherenkov photon yield is energy-dependent, understanding how energy partitioning affects the prompt photon signal is critical for optimizing timing performance in Cherenkov-based TOF-PET systems.

While these proof-of-concept studies establish a well-defined benchmark, most practical TOF-PET systems adopt pixelated crystal arrays, primarily to achieve improved spatial resolution and facilitate system integration. However, this extension to pixelated crystal arrays introduces new complexities, particularly due to Compton scattering **(Figure 1) [19-24]**. In such events, energy deposition can be distributed either across multiple crystals (inter-crystal scattering, *InterCS*) or within a single crystal (intra-crystal scattering, *IntraCS*), while the total deposited energy per detector array still falls within the photopeak window, leading to inclusion of these events in reconstructed data. Since the Cherenkov photon yield is energy-dependent, such redistribution has a direct impact on the number of prompt photons available for timing estimation.

In L(Y)SO-based detectors, both timing and energy measurements rely primarily on scintillation light, as the contribution of Cherenkov photons is negligible. When Compton scattering occurs, higher energy deposition typically takes place at interaction sites closer to the photodetector, resulting in a stronger scintillation signal and, often, earlier photon detection. This tight link between energy deposition and timing renders multiplexed readout and energy-based timestamping strategies effective in such systems.

BGO detectors, however, operate under markedly different conditions. Because fast timing relies on only a few prompt Cherenkov photons, the relationship between deposited energy and timestamp is less straightforward. Because of the low Cherenkov threshold in BGO (~63 keV), even annihilation photons undergoing maximum-energy-transfer Compton scattering (i.e. 180° backscatter) retain sufficient energy (~170 keV) to generate Cherenkov light. Consequently, in *InterCS* events, both crystals involved in the energy deposition can produce prompt Cherenkov emission, making timestamp assignment more ambiguous than in purely scintillation-based detectors. Larger energy deposits increase the overall Cherenkov yield, but *InterCS* events split energy across pixels, obscuring which timestamp best represents the event. While this decoupling does not diminish the importance of energy, it does mean that timestamp selection in BGO requires more deliberate strategies tailored to its Cherenkov photon-limited regime. Understanding and addressing these complexities is therefore essential for optimizing the timing performance of Cherenkov-based BGO TOF-PET systems.

In this work, we have systematically investigated the timing behavior of *InterCS* events in a BGO Cherenkov TOF-PET detector. Specifically, we evaluated timestamp selection strategies under realistic energy-sharing conditions, quantified changes in prompt photon yield, and analyzed timing resolution as a function of energy asymmetry. Our findings provide practical insights into the behavior of *InterCS* events and offer guidance for incorporating them into BGO TOF-PET event processing pipelines without compromising timing performance.

## 2. Materials and methods

### 2.1 Detector Configuration and Experimental Setup

The experimental setup was designed to investigate the timing behavior of *InterCS* events in a BGO-based Cherenkov TOF-PET detector. Two BGO crystal pixels, each measuring 2.5 × 2.8 × 15 mm³, were optically isolated using enhanced specular reflector (ESR) film and placed side-by-side along their 2.5 mm face. The crystals were coupled to a custom-developed OctaSiPM module, comprising eight 2.5 × 1.4 mm² SiPM channels arranged in a 2 × 4 array **[18]**. Each BGO crystal was aligned to cover two adjacent SiPM channels, enabling dual-channel readout per crystal and preserving timing sensitivity at the pixel level.

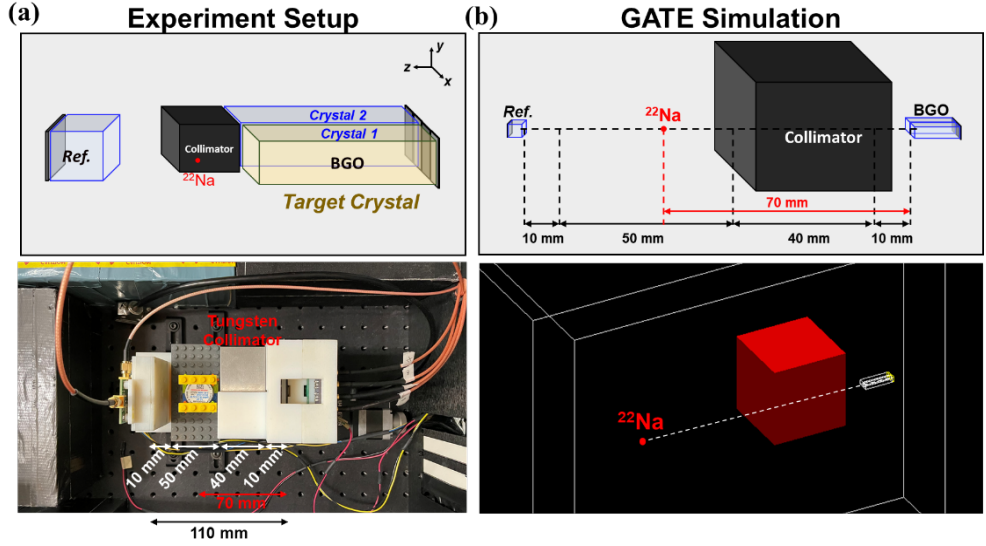

**Figure 2. Experimental setup and simulation geometry for dual-pixel BGO timing study.** (a) Experiment. (b) GATE simulation.

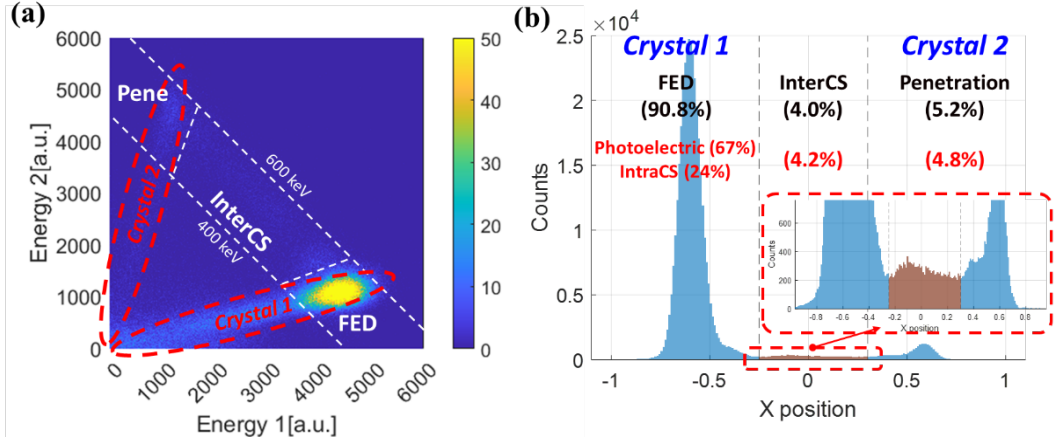

**Figure 3. Event classification via energy-weighted positioning and simulation validation.** (a) 2D histogram of deposited energy in crystal 1 and 2, in which three major clusters corresponding to *FED*, *InterCS*, and *Pene* events can be clearly seen. (b) Simulation result (red) validates the event classification via energy-weighted position profile.

A tungsten collimator was used to selectively irradiate the center of one of the two BGO crystals (designated as Crystal 1) with 511 keV annihilation photons from a $^{22}$Na point source **(Figure 2a)**. This configuration ensured that primary photoelectric interactions occurred within the targeted crystal, while allowing secondary energy deposition in the adjacent crystal via Compton scattering. Coincidence detection was performed using a reference detector based on LYSO:(Ce, Mg), which provided a fast-timing anchor for TOF measurements. All detectors operated at room temperature, and waveforms were digitized using an 8-channel high-speed oscilloscope (MSO68B, Tektronix, USA). One channel was dedicated to the reference detector, while the remaining channels recorded two energy signals (one per crystal) and four timing signals (two per crystal), allowing full capture of spatially resolved energy and timing data from the segmented BGO detector.

*2.2 Event Classification*

To investigate the timing behavior under different interaction conditions, events were classified into three categories: full-energy deposition (*FED*), *InterCS*, and penetration (*Pene*). Classification was based on the calculated interaction position derived from the energy measurements in the two BGO crystals.

The relative interaction position along the x-direction of the dual-pixel assembly was computed using the normalized energy difference:

$$\text{position} = \frac{E_2 - E_1}{E_2 + E_1}$$

where $E_1$ and $E_2$ represent the energies deposited in Crystal 1 and Crystal 2, respectively. Events with the calculated positions well within the irradiated Crystal 1 were labeled as *FED* events, while those falling predominantly within the adjacent (non-irradiated) Crystal 2 were classified as *Pene* events. Events exhibiting intermediate positions—typically spanning the boundary between the two crystals—were identified as *InterCS* events, consistent with Compton scattering followed by secondary energy deposition in the neighboring crystal **(Figure 3)**.

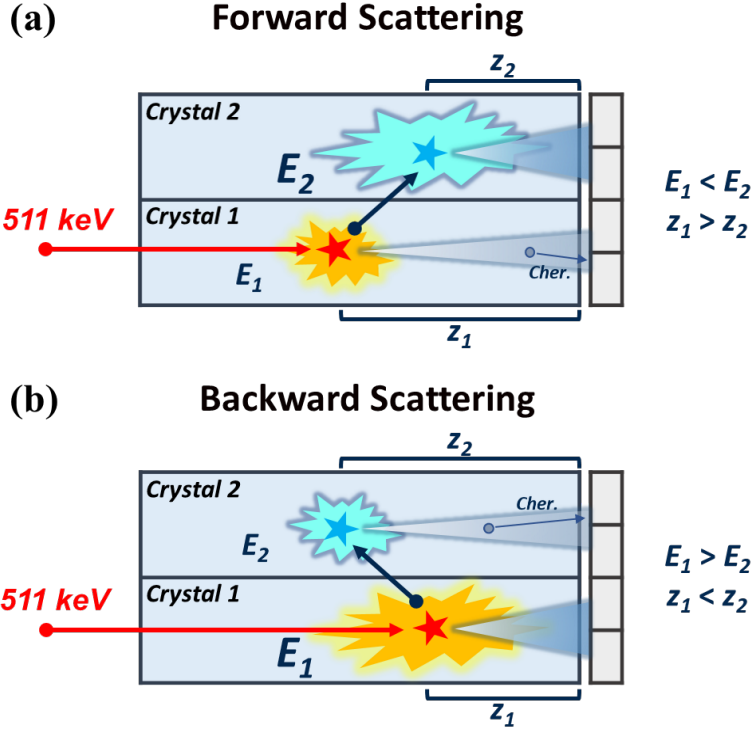

**Figure 4. Schematic illustration of energy partitioning and its geometric correlation with DOI in *InterCS* events.**
(a) Forward (FWD) scattering (b) Backward (BWD) scattering. However, both sites can still emit Cherenkov photons regardless of their relative energy, potentially decoupling energy from prompt signal generation.

To validate this classification scheme, we conducted GATE Monte Carlo simulations (vGATE 9.3) using identical detector geometry, source configuration, and collimation conditions **(Figure 2b)**. One notable advantage of the simulation is its ability to explicitly identify *IntraCS* events—i.e., interactions involving multiple energy depositions within a single crystal. Such events are are difficult to resolve experimentally due to the limited spatial granularity of the PET detectors, but may still influence the interpretation of timing behavior. The simulation revealed that a non-negligible portion of events experimentally classified as *FED* events actually involved *IntraCS* **(Figure 3b)**, underscoring the limitations of flood-map–based position estimation. To further investigate how these *IntraCS* events differ from true 511 keV *PE* events in terms of detection-time characteristics, we defined the photon travel time as the time interval between gamma-ray entrance and photon detection. We then compared the mean and standard deviation of these travel times between the two event types.

*2.3 Timestamp Selection Strategy*

For each classified event, precise timestamp selection was performed to evaluate the impact of energy distribution on timing resolution. In the case of *FED* events—where full 511 keV energy was depositied in the collimated Crystal 1—the timestamp from Crystal 1 was directly used for time pickoff. In contrast, *InterCS* events involved energy deposition across both crystals, requiring more deliberate timestamp selection strategies.

To this end, we evaluated several timestamp estimators. The primary method was the earlier timestamp selection, in which the timestamp corresponding to the earlier of the two detected signals (from Crystal 1 or Crystal 2) was chosen:

$$T_{\min} = \min(T_1, T_2)$$

where $T_1$ and $T_2$ refer to the timestamps from the Crystals 1 and 2, respectively.

To assess the robustness of this strategy, we implemented an adaptive margin-based selection rule, introducing a tunable parameter $k$ as follows:

$$T_{\text{adap}} = \begin{cases} T_1, & \text{if } T_1 < T_2 + k \\ T_2, & \text{otherwise} \end{cases}$$

By sweeping the value of $k$, we explored the trade-off between strict temporal prioritization and tolerance for signal variability. Notably, the earlier timestamp approach can be regarded as a special case of the adaptive timestamp method with a margin parameter $k=0$.

In addition to the time-based strategies, we also evaluated an energy-based adaptive timestamp selection method. When Compton scattering occurs, the incident photon depositing a larger fraction of energy typically interacts at a deeper depth—closer to the photodetector—resulting in stronger light output and an increased likelihood of earlier signal detection. This geometric correlation links energy partitioning to depth-of-interaction (DOI) and, consequently, to timing behavior. Accordingly, *InterCS* events were categorized into forward (FWD) **(Figure 4a)** and backward (BWD) **(Figure 4b)** scattering, depending on whether the irradiated crystal (Crystal 1) received less or more energy, respectively.

$$T_{\text{Energy adap}} = \begin{cases} T_1, & E_1 > E_2 \ (BWD \ scattering) \\ T_2, & E_1 \leq E_2 \ (FWD \ scattering) \end{cases}$$

This approach selects the timestamp from the crystal with the larger energy deposit, under the assumption that a higher energy interaction site is more likely to produce stronger signals and better timing precision.

While valid in scintillation-oriented detectors such as L(Y)SO—where energy and timing are closely correlated—this assumption could be less reliable in Cherenkov-based detectors like BGO, where the timing is governed by a limited number of early-arriving photons and photon detection statistics dominate the uncertainty. Because BGO has a low Cherenkov threshold (~63 keV), even the crystal receiving the smaller share of energy can still emit the Cherenkov photons, meaning that both interaction sites contribute to prompt optical emission. Therefore, comparing this energy-based adaptive approach with the earlier time-based method provides a practical way to assess how these two strategies behave under BGO's photon-limited conditions.

Each strategy was applied to the *InterCS* event group, and the resulting CTR values were compared to evaluate the effectiveness of different selection criteria under realistic energy-sharing conditions.

### 2.4 Prompt Photon Quantification

To directly evaluate how energy splitting affects early photon statistics and timing performance, we quantified the number of initially detected photons per event. Each BGO crystal was coupled to two SiPM channels (A and B), and the timing waveform from each channel was individually integrated over the first 1 ns to estimate its prompt photon contribution. The resulting photon counts from the two channels were then summed; however, if the timestamp difference between A and B exceeded a predefined coincidence tolerance window of 300 ps, the later signal was disregarded, and its photon count was set to zero. For example, if both channels individually registered two photons but the B channel timestamp lagged behind A by 400 ps, the crystal's initial photon count was recorded as two rather than four. A more detailed description of this photon count extraction method is provided in **[18]**.

For *InterCS* events—where both crystals contributed signals—the initial photon count from the crystal yielding the earlier timestamp was used. This ensured that the photon statistics corresponded specifically to the detector element that governed the final timestamp formation.

The 1 ns integration window was chosen to capture only the earliest part of the signal—where Cherenkov photons dominate timestamp formation—allowing us to isolate the component of the signal most relevant to time pickoff. This approach provides not only an indirect estimate of Cherenkov photon yield but also a direct and phenomenological indicator of how energy redistribution in *InterCS* events leads to reduced prompt photon generation.

By comparing initial photon counts across *FED*, *InterCS*, and *Pene* events, this analysis would offer direct evidence that energy sharing degrades the prompt signal component, thereby establishing a clear link energy asymmetry to measurable timing degradation.

## 3. Results

### 3.1 Classification Validation and IntraCS Fraction

Event classification based on the position profile–derived positioning metric yielded a clear separation between *FED*, *InterCS*, and *Pene* event categories. Experimentally, *FED* events dominated the dataset due to collimated irradiation on Crystal 1, while *InterCS* and *Pene* events were also consistently identified based on their intermediate and opposite-side positional signatures, respectively **(Figure 3a)**. **Figure 3a** shows three major clusters corresponding to *FED*, *InterCS*, and *Pene* events. Events near the horizontal and vertical axis correspond to Crystal 1 and 2 respectively. The oblique distribution (red dashed ellipse in **Figure 3a**) observed for each individual crystal arises from inter-crystal optical crosstalk, which occurs due to imperfect reflectance despite the use of ESR reflector wrapping **[25]**.

To evaluate the validity of this classification, a matching GATE simulation was performed using the same detector geometry and irradiation setup. The simulated distribution of event categories (written in red) showed good agreement with experimental ratios, supporting the reliability of the position profile–based approach **(Figure 3b)**. Importantly, the simulation enabled identification of *IntraCS* events, which are difficult to distinguish experimentally but can affect timing.

Among the events experimentally classified as *FED*, the simulation revealed that nearly 25% were in fact *IntraCS* events—Compton interactions with multiple energy depositions confined within a single crystal. This finding underscores the inherent limitation of spatially coarse event classification in small-scale pixelated setups and highlights the potential impact of *IntraCS* on prompt photon yield and timing variability.

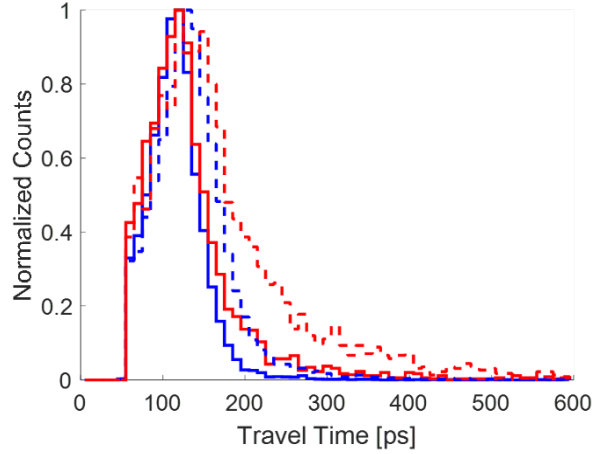

**Figure 5. Travel time profile comparison between photoelectric and *IntraCS* events from simulation.** Photon travel times for the first (n=1) and second (n=2) detected photons show delayed and broadened timing in *IntraCS* events compared to photoelectric absorption.

Indeed, in the simulation, the arrival times of detected photons were ordered for each event to examine the travel time—defined as the time interval between the gamma-ray entrance and the detection of each photon **(Figure 5)**. For the first detected photon (n=1), *IntraCS* events exhibited a significantly delayed and more dispersed distribution compared to photoelectric events, with a mean travel time of 139 ps and a standard deviation (std) of 107 ps, versus 111 ps and 36 ps for *PE* events, respectively. This clearly indicates a degradation in timing precision due to the intra-crystal scattering. The discrepancy became even more pronounced for the second detected photon (n=2), with *PE* events showing a mean of 132 ps (std 58 ps) compared to 193 ps (std 157 ps) for *IntraCS* events.

*3.2 Timestamp Difference and Adaptive Strategy*

Returning to the experimental results for *InterCS* events, the timestamp difference between Crystal 1 and Crystal 2 ($T_1-T_2$) exhibited a symmetric distribution centered around zero, spanning approximately ±3 ns **(Figure 6a)**. This indicates that neither crystal consistently leads in time across events, and that the timing behavior is dominated by stochastic variations in photon production and transport, rather than any systematic asymmetry in geometry or electronics.

To determine the optimal strategy for timestamp selection in *InterCS* scenarios, we evaluated the performance of an adaptive timestamp rule using a tunable margin parameter $k$. As defined in Section 2.3, the earlier of $T_1$ and $T_2$ was selected if $T_1 < T_2 + k$; otherwise, $T_2$ was used. Sweeping $k$ across a range of values revealed that the best timing resolution was achieved when $k=0$, i.e., when the strictly earlier timestamp was always selected **(Figure 6c)**. Although the CTR for *InterCS* events at $k = 0$ was measured to be 221 ps FWHM, noticeably worse than the 184 ps FWHM observed for *FED* events **(Figure 6b)**. This indicates that, despite optimizing the timestamp rule, the intrinsic nature of energy splitting in *InterCS* events leads to fewer prompt photons and increased temporal uncertainty, ultimately degrading timing performance relative to full-energy deposition cases.

We also compared this time-based strategy with an energy-based timestamp selection approach, where the timestamp from the crystal with the larger energy deposit was chosen. While this method is often effective in scintillation-dominated materials such as L(Y)SO—where a larger energy deposit typically indicates an interaction occurring closer to the photodetector and thus earlier photon arrival—it performed suboptimally in our BGO Cherenkov detector. A significant fraction of *InterCS* events showed no correspondence between energy dominance and timestamp order, highlighting the decoupling of energy and timing in photon-starved Cherenkov conditions **(Figure 6d)**.

Overall, the results confirm that a simple first-hit timestamp strategy ($k=0$) outperforms energy-based selection for *InterCS* events in BGO detectors, and provides the most consistent timing performance across a wide range of energy sharing scenarios.

*3.3 Timing vs. Energy Asymmetry*

To further understand the origin of timing variability in *InterCS* events, we analyzed the timing resolution as a function of energy asymmetry between the two crystals. In the simple binary classification into FWD ($E1>E2$) and BWD ($E1<E2$) scattering, as shown in **Figure 7a**, the distribution was nearly symmetrical, with a slightly higher fraction of FWD events, indicating that both scattering types occur at comparable rates under the collimated irradiation setup. Beyond the binary classification, we computed a continuous energy asymmetry metric and binned the events accordingly to examine resolution trends across varying degrees of energy imbalance. As shown in **Figure 7b**, events with highly asymmetric energy deposition consistently exhibited better timing resolution than those with nearly equal energy sharing.

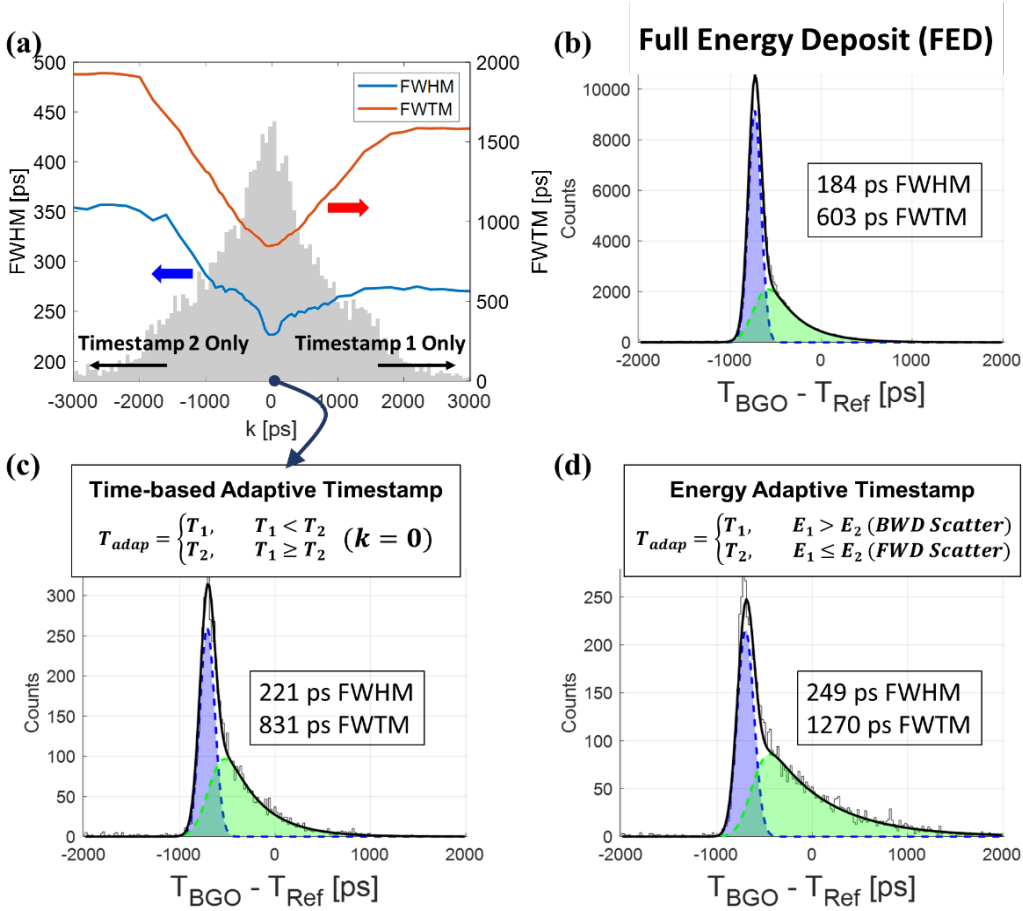

**Figure 6. Timestamp difference distribution and timing performance under adaptive selection.** (a) Histogram of timestamp differences between the two crystals for *InterCS* events. (b) *FED* events timing histogram. (c) *InterCS* events timing histogram at optimal *k*=0 (d) *InterCS* events timing histogram with energy adaptive timestamp.

This observation can be attributed to the higher Cherenkov photon yield in the crystal receiving more energy, which leads to a stronger and more prompt signal for timestamp formation **(Figure 7c)**. In contrast, when energy is more evenly split across both crystals, neither channel provides a sufficiently strong early-photon signal, resulting in degraded timing precision. These results confirm that energy sharing negatively impacts early photon generation—and that the worst timing occurs when energy is divided nearly equally between the two crystals.

*3.4 Prompt Photon Yield Analysis*

To directly assess how energy splitting in *InterCS* events affects early photon availability, we compared the number of promptly detected photons across the three event categories. The prompt photon count was estimated by integrating the first 1 ns of the SiPM timing signal waveform from the crystal corresponding to the earlier timestamp, as described in Section 2.4.

As shown in **Figure 8a**, *InterCS* events exhibited a lower average prompt photon count (4.73 photons) compared to *FED* events (5.76 photons), indicating a clear reduction in prompt photon generation due to energy redistribution. In contrast, *FED* and *Pene* events showed comparable mean photon counts (5.76 and 5.33 photons), suggesting that full energy deposition—regardless of crystal location—leads to a similar prompt photon yield **(Figure 8b)**. The slight reduction in the photon yield for *Pene* events compared to *FED* events may be attributed to a small amount of energy loss due to scattering at the collimator edge or to the fact that penetration events tend to occur near crystal edges where photon collection is less efficient. Although the absolute difference in photon count between *FED* and *InterCS* events is just over one photon, its impact is non-negligible in the context of BGO Cherenkov detectors, where timing estimation relies on only a few early-arriving photons. This result provides direct evidence that energy splitting in *InterCS* events degrades the prompt signal component most critical for accurate timestamping, offering a mechanistic explanation for the timing resolution loss observed in Section 3.2.

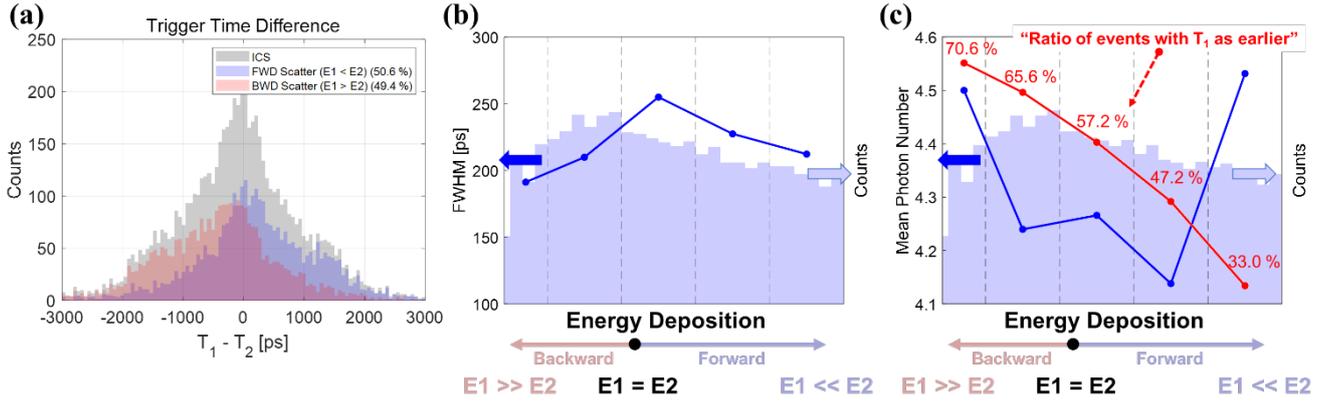

**Figure 7. Energy asymmetry in *InterCS* events.** (a) Distribution of forward and backward scatter events. (b) CTR trend according to energy deposition asymmetry. (c) Initial photon number according to energy deposition asymmetry. When viewed together with the ratio of events having $T_1$ as the earlier timestamp (red line), it becomes evident that large energy asymmetry leads to a stronger Cherenkov yield in the higher-energy crystal, producing a more prompt signal for timestamp formation.

## 4. Discussion

### 4.1 Why InterCS Matters in Cherenkov-Based BGO Systems

*InterCS* events are an inherent feature of pixelated PET detector arrays, arising when a 511 keV annihilation photon undergoes Compton scattering in one crystal and deposits its remaining energy in an adjacent crystal. While in single-crystal setups these scattered events can often be excluded by energy gating, in realistic PET systems with segmented arrays, they frequently remain within the photopeak energy window and are thus included in reconstruction data to maintain system sensitivity and efficiency.

In the context of Cherenkov-based timing with BGO, *InterCS* events introduce particular challenges that are more severe than in traditional scintillation-based detectors. Unlike L(Y)SO scintillators—which produce abundant photons with time profiles that maintain tight coupling between energy deposition and signal timing—BGO relies on the detection of just a few early-arriving Cherenkov photons for timing pickoff. The Cherenkov photon yield is roughly proportional to the deposited energy. Therefore, when energy is split between crystals during an *InterCS* event, the prompt photon yield at each site is reduced. This energy redistribution can directly limit the number of detectable early photons available for timestamp formation, degrading timing resolution.

Moreover, the stochastic nature of Compton scattering paths means that energy sharing between crystals can vary widely, from highly asymmetric to nearly equal partitioning. Our analysis showed that events with balanced energy splitting exhibit significantly worse timing performance, since neither crystal has sufficient energy deposition to produce a robust early-photon signal. Conversely, highly asymmetric *InterCS* events can maintain acceptable timing because one crystal still receives enough energy to yield Cherenkov photons capable of precise timestamping.

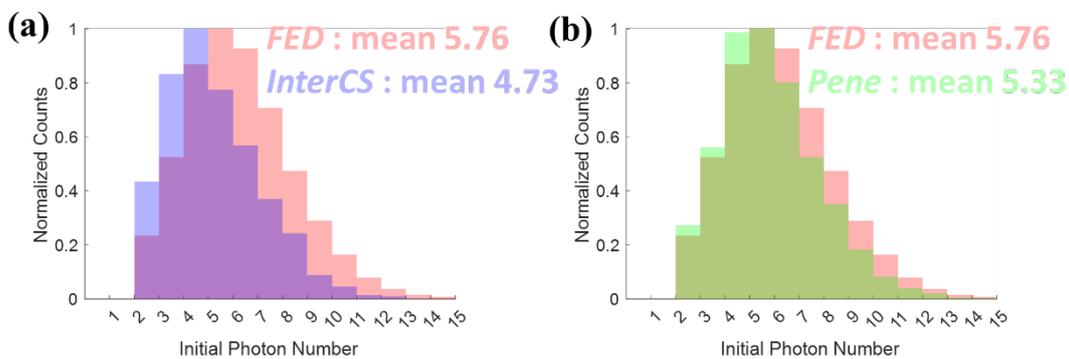

**Figure 8. Initial photon number within the first 1 ns comparison across *FED*, *InterCS*, and *Pene* events.** (a) *FED* vs. *InterCS*. (b) *FED* vs. *Pene*.

Recognizing these effects is essential for accurate modeling and optimization of TOF-PET system performance. Simply applying event selection strategies developed for scintillation-based systems—such as energy-based timestamp choice—proves insufficient for BGO Cherenkov detectors, given the decoupling of energy magnitude and photon arrival timing in photon-starved conditions. By explicitly analyzing *InterCS* events and quantifying their impact on early photon yield and timing resolution, this study provides important evidence that effective event-by-event timestamp strategies should take into account the unique physics of Cherenkov emission in BGO. Addressing these challenges is vital to fully exploiting BGO's potential as a hybrid Cherenkov/scintillation detector in practical TOF-PET systems.

*4.2 Comparison with Energy-based Timing Logic*

Traditional TOF-PET systems employing L(Y)SO crystals rely on scintillation processes that produce abundant photons with well-defined time profiles. In these detectors, the time pickoff is typically derived from multiplexed readout channels that aggregate signals across multiple pixels, and the timestamp is often selected based on the crystal with the largest measured energy deposit **[26]**. This strategy is effective because in L(Y)SO-based scintillation-oriented systems, the energy deposition tends to correlate with DOI **(Figure 9a)**. As a result, the crystal receiving more energy is generally the one closer to the interaction site, yielding a faster and cleaner timing signal. This tight coupling between energy magnitude and photon arrival time underpins the success of energy-weighted timestamp selection in these systems.

However, Cherenkov-based BGO detectors operate under fundamentally different conditions. Cherenkov emission in BGO is photon-starved, typically yielding only a handful of prompt photons per event. These early photons are critical for achieving precise timing resolution, but their production and detection are dominated by stochastic factors such as emission angle, photon path length, and local energy deposition. Unlike scintillation, Cherenkov photon production is not characterized by a large, smooth light pulse that reliably preserves the timing structure of energy deposition.

Our experimental results confirm that in BGO-based systems, the crystal with the larger energy deposit in *InterCS* events does not consistently yield the earlier timestamp **(Figure 9b,c)**. We observed that a significant fraction of events showed a mismatch between energy dominance and timestamp order, demonstrating a decoupling of energy amplitude from photon arrival time. This means that energy-based timestamp selection, which works well in L(Y)SO, is suboptimal in BGO Cherenkov systems.

Instead, the best timing performance in our study was achieved by adopting a time-based, event-by-event selection strategy that simply chose the earliest detected timestamp regardless of energy. This approach leverages the actual arrival dynamics of the sparse prompt photons rather than assuming that energy magnitude predicts timing quality. The contrast between the two materials underscores the importance of detector-specific timestamp strategies: methods effective in scintillation-based detectors may not be directly applicable to Cherenkov-based BGO systems without risking significant timing degradation.

It is important to interpret these results carefully. While the event-by-event decoupling between energy magnitude and timestamp ordering indeed increases, larger energy deposits still tend to generate more Cherenkov photons, thereby providing, on average, a faster and more stable timing estimates. Therefore, the superiority of the first-hit-based timestamp selection shown in **Figure 6** is fully consistent with the trends observed in the energy asymmetry analysis of **Figure 7**.

*4.3 Implications for Detector Readout Architecture (Per-Pixel vs. Multiplexed)*

The findings of this study provide useful considerations for the choice of readout architectures in Cherenkov-based BGO TOF-PET detectors. In conventional scintillation-based systems such as those using L(Y)SO, multiplexed readout is commonly employed to simplify electronics, reduce channel count, and combine timing signals from multiple crystals **[27]**. This approach can be highly effective because the large scintillation photon yield per event produces strong, well-defined pulses, allowing energy-weighted methods to reliably identify the earliest interaction site. As a result, multiplexed timing channels can often maintain excellent performance even in complex detector blocks.

In Cherenkov-based BGO systems, the situation differs primarily due to the smaller number of prompt photons available for timing estimation. These early-arriving photons are more susceptible to variations in photon transport paths, optical interfaces, and surface conditions. In cases where inter-crystal scatter occurs, energy splitting between pixels further reduces the prompt photon yield in each pixel. Preserving distinct per-pixel timing information can therefore be advantageous, as it allows the timing algorithm to adapt to the statistical fluctuations inherent to Cherenkov photon detection.

Our results show that event-by-event, per-pixel timestamp selection—particularly the first-hit method—provided consistent performance advantages over energy-weighted approaches under the tested conditions. This suggests that readout architectures offering independent high-resolution timing for each pixel may help retain the full timing potential of BGO in Cherenkov-based TOF-PET, even if they require more channels than multiplexed designs. Such architectural choices should be considered in balance with other system-level constraints, such as complexity, cost, and integration requirements.

*4.4 Considerations for IntraCS Events*

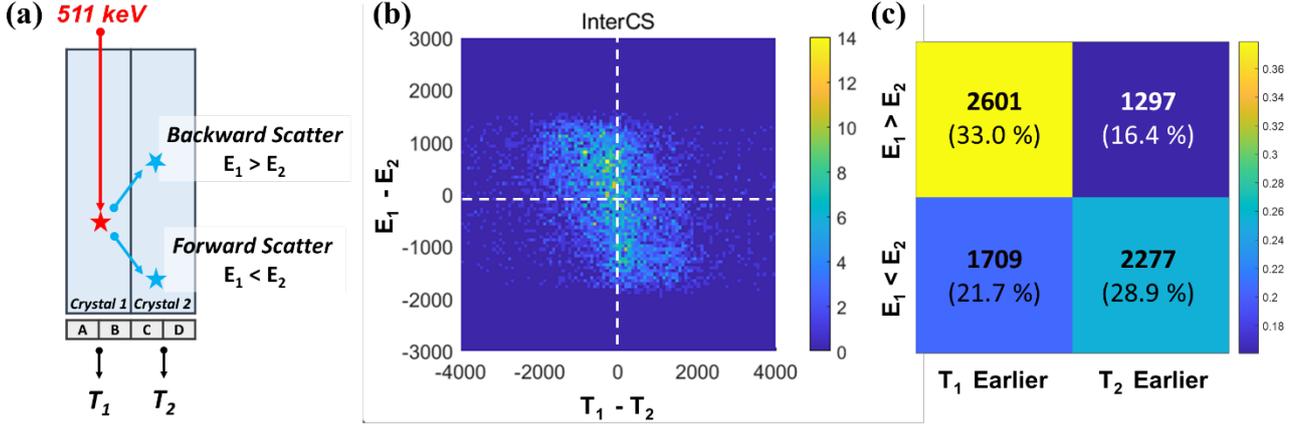

**Figure 9. Decoupling of energy dominance and timestamp order in BGO *InterCS* events.** Demonstrates why energy-based timestamp selection fails in photon-starved BGO detectors. (a) Forward and backward scatter. (b) Relationship between energy asymmetry ($E_1$ vs. $E_2$) and timestamp ordering ($T_1$ vs. $T_2$) in *InterCS* events. (c) Each quadrant represents a distinct combination of energy and timing order, and the relative fraction of events falling into each region is annotated to highlight the decoupling between energy dominance and signal arrival time.

The *IntraCS* events also represent an important and interesting category to consider when evaluating timing performance in BGO-based Cherenkov TOF-PET systems. By definition, *IntraCS* involves multiple energy depositions within a single crystal, making it practically indistinguishable from simple photoelectric events when relying on coarse positional granularity. However, as our simulation results have shown, the Cherenkov photon yield is highly sensitive to deposited energy. This sensitivity means that *IntraCS* events can have a substantial and measurable impact on timing resolution in Cherenkov-based applications.

Our previous study demonstrated that enhancing direct detection efficiency by properly suppressing reflections of Cherenkov photons can maximize the usage of their promptness [16]. One strategy to achieve this is to reduce the crystal aspect ratio, even approaching a monolithic configuration. However, it is important to recognize that such adjustments may increase the proportion of *IntraCS* events. Indeed, one simple simulation exploring different pixel geometries (with fixed 15 mm length but varying transverse pitches of 1, 2, 3, 5, 10, and 15 mm) **(Figure 10a)** revealed that as the aspect ratio decreases, the fraction of events involving multiple energy depositions within a single crystal increases **(Figure 10b)**. This indicates that while reducing reflections improves direct prompt photon collection, it also increases the likelihood of *IntraCS* interactions, even approaching 50% with the monolithic configuration (15 × 15 × 15 mm$^3$), which can degrade timing performance by distributing energy across multiple positions within the same crystal.

Additionally, beyond timing considerations, *IntraCS* events may also influence DOI estimation. In designs employing dual-ended readout or light-sharing schemes that estimate DOI from energy ratios, multiple-position energy deposition within a single crystal can introduce DOI estimation errors. However, our simulation analysis suggests that this effect may be limited. Specifically, when examining the subset of *IntraCS* events in simulation, the proportion showing a DOI separation (ΔDOI) greater than 2 mm between the first Compton scattering point and the final photoelectric absorption was not particularly large. This suggests that, unlike timing performance—which is highly sensitive to prompt photon statistics—the impact of *IntraCS* on DOI estimation may not be as significant in practice.

This distinction may also explain why the *IntraCS* events have historically received less attention in traditional lutetium-based scintillation detectors. In those systems, the detected signal effectively represents the sum of all scintillation light generated within the crystal, so there has been little practical need to treat them as a separate category from single-site photoelectric interactions. As a result, the impact of *IntraCS* on timing or DOI estimation has generally been considered negligible. In contrast, the photon-starved nature of Cherenkov-based BGO timing requires careful attention to even subtle effects like *IntraCS* to optimize detector performance.

## 5. Conclusions

This study systematically investigated the impact of Compton scattering on the timing performance of BGO-based Cherenkov TOF-PET detectors using a dual-pixel setup with segmented SiPM readout. We found that *InterCS* events, which are inevitable in pixelated detector arrays, introduce notable timing degradation due to energy redistribution between adjacent crystals. This redistribution reduces the number of early-arriving Cherenkov photons, with *InterCS* events yielding approximately one fewer photon on average in the first 1 ns compared to events with full 511 keV deposited. Critically, our analysis revealed that timestamp selection strategies based on energy dominance are suboptimal in photon-starved BGO environments. Instead, a simple event-wise selection of the earliest timestamp outperformed energy-based approaches. Furthermore, Monte Carlo simulations highlighted the non-negligible presence of intra-crystal scattering and its associated timing uncertainty. These findings emphasize the importance of photon-aware, per-pixel timestamping architectures and event-level considerations in BGO Cherenkov detector design, and provide practical insights for optimizing BGO Cherenkov TOF PET system performance.

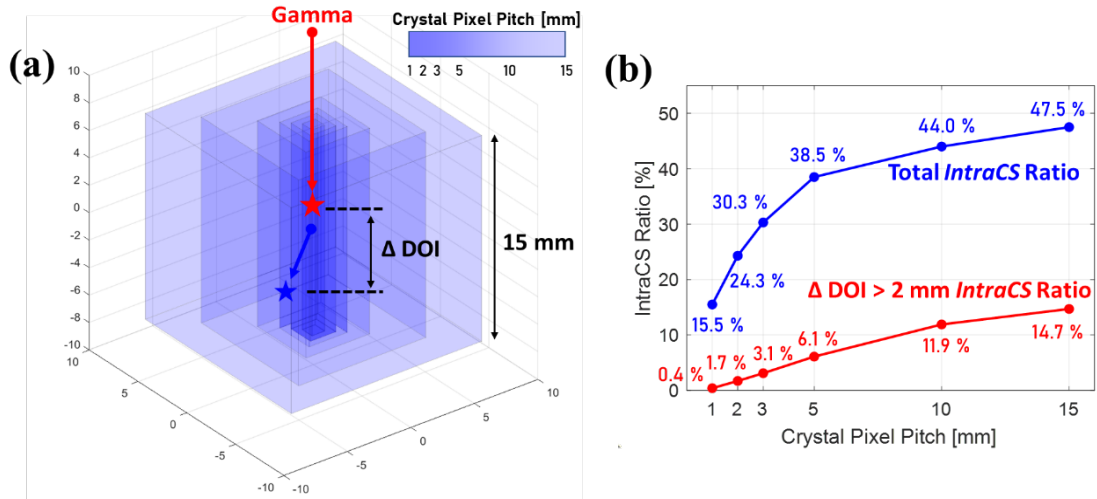

**Figure 10. Effect of crystal aspect ratio on *IntraCS* prevalence and timing considerations.** (a) Pixel configurations with varying transverse sizes. (b) Fraction of *IntraCS* events increases with decreasing aspect ratio, with implications for increasing timing contribution.

## Acknowledgments

This work was supported by a National Institutes of Health grant R01 EB029633 and the National Research Foundation of Korea (NRF) grant funded by the Korea government (MSIT) (Grant No. RS-2024-00354123).